\documentclass[twocolumn,preprintnumbers,amsmath,amssymb]{revtex4}
\usepackage{color}
\usepackage{graphics}
\usepackage{dcolumn}
\usepackage{bm}

\renewcommand\textfraction 0
\renewcommand\topfraction 1
\renewcommand\bottomfraction 1
\begin{document} 

\preprint{Applied Physics A (2004) accepted}

\title{Silicon clusters produced by femtosecond laser ablation: Non-thermal emission and gas-phase condensation}
\author{Alexander V. \surname{Bulgakov}}
\altaffiliation{Fax: +7 3832 343480, Tel.: +7 3832 391045}
\email{bulgakov@itp.nsc.ru}
\affiliation{Institute of Thermophysics, Prospect Lavrentyev 1, 630090 Novosibirsk, Russia}
\author{Igor \surname{Ozerov}}
\email{ozerov@crmcn.univ-mrs.fr}
\affiliation{CRMC-N, UPR 7251 CNRS, Universit\'{e} de la M\'{e}diterran\'{e}e, Facult\'{e} des Sciences de Luminy, Case 901, 13288 Marseille, Cedex 9, France}
\author{Wladimir \surname{Marine}}
\email{marine@crmcn.univ-mrs.fr}
\affiliation{CRMC-N, UPR 7251 CNRS, Universit\'{e} de la M\'{e}diterran\'{e}e, Facult\'{e} des Sciences de Luminy, Case 901, 13288 Marseille, Cedex 9, France}
\date{\today}

\begin{abstract}
Neutral silicon clusters Si$_{n}$ (up to $n = 7$) and their 
cations Si$_{n}^{ + }$ (up to $n = 10$) have been produced by femtosecond 
laser ablation of bulk silicon in vacuum and investigated using 
time-of-flight mass spectrometry. Two populations of the Si$_{n}^{ + }$ 
clusters with different velocity and abundance distributions in the ablation 
plume have been clearly distinguished. Possible mechanisms of cluster 
formation (Coulomb explosion, gas-phase condensation, phase explosion) are 
discussed.

\end{abstract}

\maketitle

Femtosecond laser ablation is a rapidly developing technique offering new 
possibilities in various applications. The fundamental mechanisms that lead 
to material removal are, however, still poorly understood. The interaction 
of fs laser pulses with silicon is an example of a complex interplay of 
thermal and ultrafast, non-thermal processes involved to ablation \cite{1,2,3}. 
Studies of composition and expansion dynamics of the laser-induced plume can 
provide a considerable insight into the ablation mechanisms. Several recent 
experiments on this subject \cite{2,4,5,6} were mainly devoted to the atomic 
component of the plume though an observation of small silicon clusters was 
also mentioned \cite{2}. Little is known about silicon cluster formation under 
ablation with short laser pulses. With ns pulses, low-fluence desorption of 
Si dimers \cite{7} and small neutral Si$_{n}$ clusters \cite{8} were observed. 
Desorption of small clusters from both crystalline and nanostructured Si 
surfaces was induced by using high-energy (6.4 eV) photons \cite{9}. 
Multiple-charged cluster ions were formed in ps-laser stimulated field 
evaporation \cite{10}. Recently we reported the first results on the observation 
of neutral and cationic clusters under femtosecond laser ablation of silicon 
\cite{11}. In this work, we analyze mechanisms of clusters formation based on 
measurements of the abundance and velocity distributions.

\section{Experiment}

The experiments were performed with Si[100] surface under ultrahigh vacuum 
conditions ($\sim $ 10$^{ - 10}$ mbar). The Si target was irradiated at a 
45\r{ } incidence angle using a Ti:sapphire laser (Mai-Tai coupled with a 
TSA amplifier, Spectra Physics, 80 fs pulse duration, 10 Hz repetition rate, 
up to 30 mJ energy per pulse) operating at 800 nm. A part of the laser beam 
was selected by an aperture to provide a nearly uniform intensity 
distribution over the irradiated spot. The target was rotated/translated 
during measurements to avoid considerable cratering. Some experiments were 
performed with the fixed target in order to investigate the effect of 
accumulation of laser pulses at the same spot. The fluence on the target was 
varied in the range 80-800 mJ/cm$^{2}$. 

The abundance distributions of neutral and positively charged particles in 
the laser-ablation plume were studied using a reflectron time-of-flight mass 
spectrometer. The neutral particles were analyzed using electron impact 
ionization (110 eV) and a plasma suppressor \cite{12}. At a distance of 11 cm 
from the target, the ablated or post-ionized ions were sampled parallel to 
the plume axis by a 500 V repeller pulse at a time delay $t_{d}$ in respect 
to the laser pulse. Mass spectra were averaged over 300 laser shots with the 
rotated target and over 5 shots with the fixed target.

\section{Results and discussion}

\begin{figure}[!]
\begin{center}
\resizebox{9cm}{!}{\includegraphics{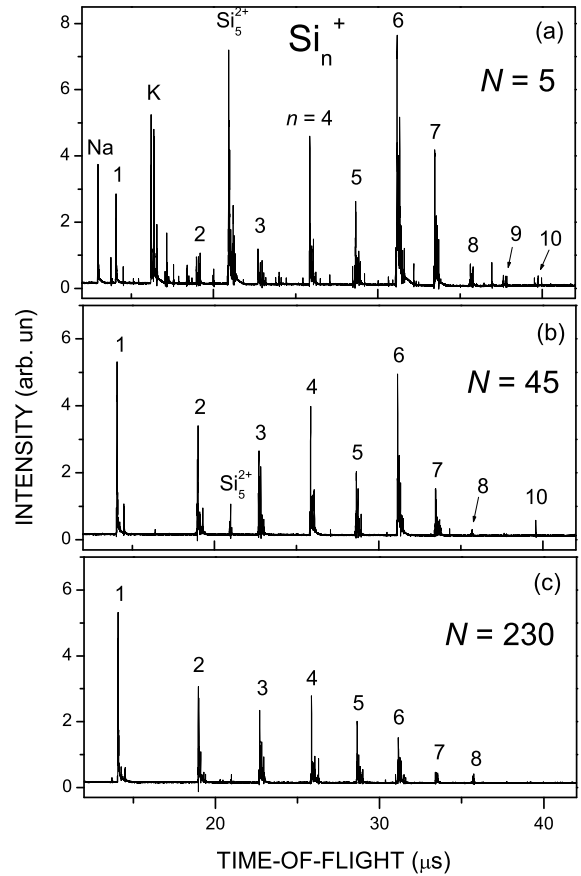}}
\end{center}
\caption{\label{fig1}
Mass spectra of cationic silicon clusters produced at laser 
fluence $F$ = 350 mJ/cm$^{2}$ with different number $N$ of laser pulses applied. 
Time delay $t_{d}$ = 32 $\mu $s}
\end{figure}

Monatomic Si$^{ + }$ ions and Si atoms are the most abundant particles in 
the plume throughout the laser fluence range studied. The ion signal appears 
at a threshold fluence $F_{th} \approx$ 100 mJ/cm$^{2}$ \cite{11} for the 
etched Si surface irradiated previously with a fairly large number of laser 
pulses. Efficient emission of both neutral silicon clusters Si$_{n}$ (up to 
$n$ = 7) and their cations Si$_{n}^{ + }$ (up to $n$ = 10) has been observed for 
$F > F_{th}$. The cluster abundance distributions are found to be essentially 
different for the fresh and etched ablated surfaces. Figure \ref{fig1} shows a 
typical evolution of cation mass spectra with increasing the number $N$ of 
laser pulses applied to the same spot. The spectra were obtained at time 
delay $t_{d}$, corresponding to a maximum average yield of clusters. With the 
fresh surface, the relatively large Si$_{n}^{ + }$ clusters ($n > 3$) are 
observed in high abundance with the "magic" number at $n$ = 6 (Fig. \ref{fig1}a). In 
addition, doubly charged cluster ions are present in the plume among which 
Si$_{5}^{2 + }$ is particularly abundant (note that Si$_{5}^{2 + }$ was 
observed earlier as a magic cluster in the field evaporation experiment 
\cite{10}). The monatomic Si$^{ + }$ peak is relatively weak under these 
conditions.

When the number of laser shots increases, the mass spectrum is significantly 
modified (Fig. \ref{fig1}b,c). The Si$_{n}^{ + }$ ($n > 3$) and Si$_{5}^{2 + }$ 
clusters are observed in lower abundance (the latter almost disappears at 
$N > 100$) while the concentration of the smaller species ($n$ = 1-3) increases 
progressively with $N$. After $\sim $40 laser shots, Si$^{ + }$ ion is totally 
dominant in the plume (this is not seen in Fig. \ref{fig1} since the Si$^{ + }$ signal 
is maximized at shorter time delays). The cluster distribution exhibits an 
odd-even alternation with preferred formation of even-numbered Si$_{n}^{ + 
}$ clusters. Further increase in the number of laser pulses results in 
further decrease of the relative yield of the cationic clusters. After few 
hundreds of laser shots the signal saturates and the abundance distribution 
becomes rather smooth with peak intensities monotonously decreasing with 
cluster size (Fig. \ref{fig1}c).

\begin{figure}[!]
\begin{center}
\resizebox{8.5cm}{!}{\includegraphics{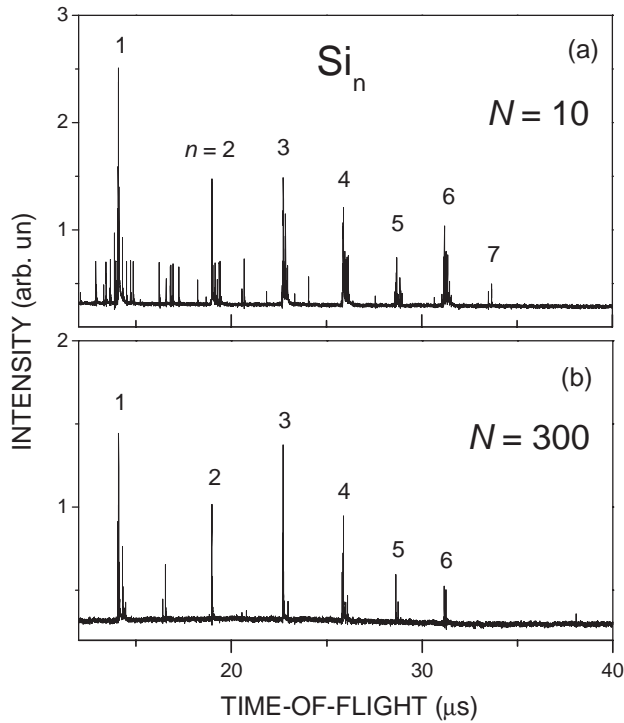}}
\end{center}
\caption{\label{fig2}
Mass spectra of neutral silicon clusters produced at $F$ = 350 
mJ/cm$^{2}$ and detected at $t_{d}$ = 32 $\mu $s with relatively fresh (a) 
and etched (b) surfaces.}
\end{figure}

Typical mass spectra for neutral Si$_{n}$ clusters obtained with the fresh 
and etched ablated surfaces are shown in Fig. \ref{fig2}. The effect of the number of 
accumulated laser pulses is not so strong in this case though the same 
tendency of decreasing in the relative yield for Si$_{n}$ clusters ($n > 4$) is 
obvious. The odd-even alternation in the cluster distribution is also 
observed at low $N$ values (Fig. \ref{fig2}a) but it is less pronounced than for 
cationic clusters.

The observed effect of the number of applied laser pulses on particle 
emission suggests an accumulation of the laser-induced damage on the Si 
surface. Indeed, with an optical microscope we observed clearly shaped spots 
at the surface for very low fluences down to approximately 150 mJ/cm$^{2}$ 
(for multi-pulse conditions with $\sim $1000 laser shots per spot). We can 
thus conclude that the damage threshold is approximately equal to the ion 
desorption threshold for silicon under these conditions. The low-fluence 
damage (modification) of Si surfaces with ultrashort laser pulses was 
observed earlier and attributed to amorphization of silicon \cite{13,14}. The 
effect of laser pulse accumulation on the damage threshold (the incubation 
effect) was also revealed. The multi-pulse damage threshold determined in 
\cite{14} is well consistent with that evaluated here. Similar incubation effect 
in silicon was observed recently with ns laser pulses \cite{8,15}. The precise 
nature of the incubation effect is not clear (it can be, e.g., due to 
chemical modification, mechanical stress, defect accumulation \cite{14,15}) and 
needs further investigation.

\begin{figure}[!]
\begin{center}
\resizebox{9cm}{!}{\includegraphics{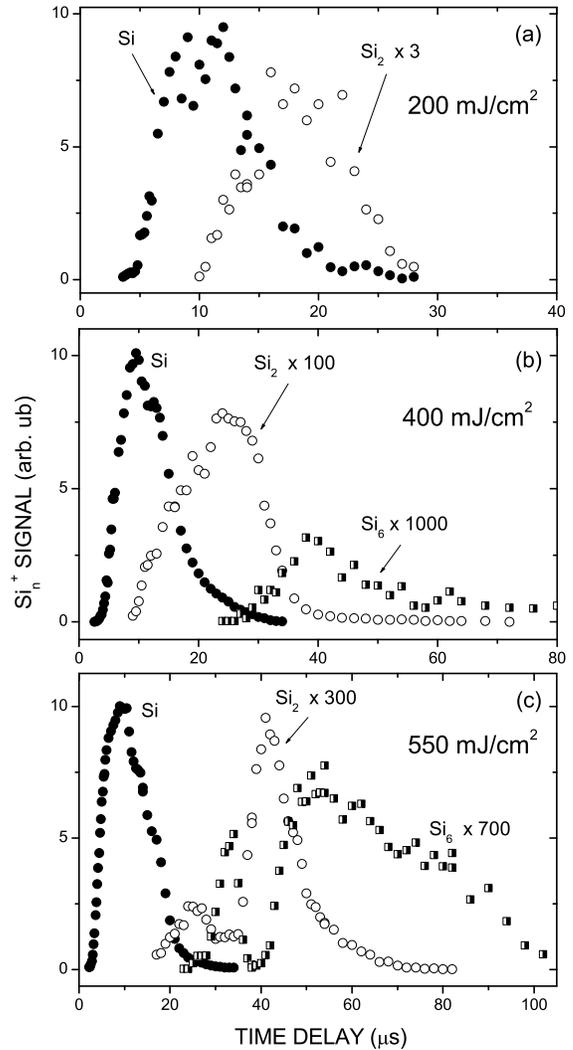}}
\end{center}
\caption{\label{fig3}
TOF distributions for Si$^{ + }$, Si$_{2}^{ + }$ and 
Si$_{6}^{ + }$ cations at three laser fluences.}
\end{figure}

Variation of the time delay $t_{d}$ allowed the analysis of particle velocity 
distributions and the characterization of the temporal evolution of the 
plume composition. The measurements were perfermed with the etched Si 
surface when no signal change with the number of laser pulses occurred. Figure 
\ref{fig3} shows the time-of-flight (TOF) distributions of several cationic particles 
for different laser fluences. The distributions for Si$^{ + }$ ions are 
maximized at $\sim $9 $\mu $s time delay that corresponds to $\sim $12 km/s 
ion velocity. Increase in fluence results just in a broadening of the 
distributions with little or no shift towards higher velocities. At the same 
time, the yield of Si$^{ + }$ ions increases strongly with fluence as $\sim
F^{n}$ with $n \approx $ 7.5 \cite{11}.

In contrast to Si$^{ + }$ ions, the cluster TOF distributions are found to 
be strongly fluence-dependent. At low fluence, near the threshold value, the 
TOF distribution for Si$_{2}^{ + }$ is rather narrow, single-peaked, and 
maximized at $t_{d} \cong $ 18 $\mu $s (Fig. \ref{fig3}a), that is the most 
probable velocity is approximately two times lower than that of Si$^{ + }$. 
This means that these plume particles have nearly the same momentum. At 400 
mJ/cm$^{2}$, the distribution for Si$_{2}^{ + }$ is still single-peaked 
but a noticeable shift towards higher time delays is observed (the maximum 
occurs at $\sim $25 $\mu $s, the corresponding velocity is $\sim $ 4.4 
km/s). The fast Si$_{2}^{ + }$ dimers are still present in this ablation 
regime but their distribution is masked by slower ions. With further 
increase in laser fluence, the second, even slower, population of 
Si$_{2}^{ + }$ appears in the plume. At 550 mJ/cm$^{2}$ this slow 
population dominates (Fig. \ref{fig3}c). The first faster peak in the distribution is 
still present, well separated from the second peak, and is still maximized 
at $\sim $25 $\mu $s. 

The evolution of TOF distributions for larger Si$_{n}^{ + }$ clusters is 
qualitatively similar to that for Si$_{2}^{ + }$ as illustrated in Fig. \ref{fig3} 
for the S$_{6}^{ + }$ cation. At a threshold fluence of around 450 
mJ/cm$^{2}$, the distributions are transformed from single-peaked to 
double-peaked. The second (slow) cluster population becomes rapidly dominat 
with further increase in fluence. The most probable velocity of the second 
population decreases slightly with cluster size from 2.5 km/s for 
Si$_{2}^{ + }$ to 2 km/s for Si$_{6}^{ + }$. The abundance distribution 
for this slow cluster population exhibits a pronounced odd-even alternation 
with dominance of the Si$_{6}^{ + }$ peak and similar to that observed at 
shorter delay times with relatively fresh surface (Fig. \ref{fig1}a). It is thus 
remarkably different from the falling smooth distribution for faster 
clusters (Fig. \ref{fig1}c). 

Based on the obtained results, we can address the fundamental question on 
cluster formation mechanism under different irradiation regimes. At very low 
laser fluence, from the ion appearance threshold $F_{th}$ up to $\sim $2$ F_{th}$, 
Si$^{ + }$ and Si$_{2}^{ + }$ ions are likely produced by an impulsive 
Coulomb explosion (CE) from a charged surface. Two observations support the 
CE mechanism \cite{11,16}, namely, (a) the momentum scaling for the particles 
under low fluence regimes and (b) decrease of velocity for Si$_{2}^{ + }$ 
ions when fluence is increased beyond $\sim $2$F_{th}$, that is a behavior 
opposite to thermal desorption. The slower Si$_{n}^{ + }$ clusters (which 
overshadow the CE ions in the (2-4)$ F_{th}$ fluence range and form the fast 
cluster population at higher fluences) can be interpreted as "plasma ions", 
i.e., ions formed in the ionized vapor plume by collision-induced 
condensation. The velocities of these "plasma ions" decrease only slightly 
with cluster size and scale by a law intermediate between constant kinetic 
energy and constant velocity that implies the gas-phase condensation 
mechanism rather than direct ejection of the clusters \cite{12}. 

Of particular interest is the origin of very slow cluster ions observed at 
high fluences ($F > 4.5 F_{th})$ as the second population in the TOF 
distributions (Fig. \ref{fig3}c). The evident separation of this population from the 
first one indicate that another mechanism is likely responsible for its 
formation. We suggest that the slow ions are due to phase explosion (PE) 
from the Si surface melted via an ultrafast, non-thermal process which was 
recently found for semiconductors under fs-laser irradiation \cite{1,2,3}. The fast 
excitation of a dense electron-hole plasma causes destabilization of the 
lattice, and the semiconductor melts in a time less than 1ps. The 
non-thermal melting occurs at fluences fairly higher than needed for purely 
thermal melting process. The melted surface, even if its temperature is far 
below the thermodynamic critical temperature, inevitably enters the region 
of metastable states with the subsequent phase explosion as one of probable 
stabilization mechanisms \cite{17}. The overheated liquid can decay into the gas 
phase without formation of the critical vapor nucleus \cite{17} so the ejection 
of small clusters is quite possible. More work is needed to be certain about 
the origin of the cluster ions under these conditions but some observations 
support the PE mechanism. The amount of material released under the PE is 
macroscopic and the ejected particles undergo many collisions within the 
plume during their expansion into vacuum. As a result, the most stable 
clusters survive predominantly in the plume thus resulting in abundance 
distributions with magic numbers. Numerous previous investigations indicate 
special stability of Si$_{4}^{ + }$, Si$_{7}^{ + }$, and particularly 
Si$_{6}^{ + }$ \cite{18}. This is well correlated with the observed abundance 
distributions of the slow ions (similar to that shown in Fig. \ref{fig1}a). In 
addition, the observed weak size dependence of cluster velocities for the 
second population (Fig. \ref{fig3}c) correlate with recent simulations of short-pulse 
laser ablation \cite{19} where nearly the same flow velocities were obtained for 
clusters of different size ejected via the PE mechanism.

\section{Conclusions}

We have observed efficient emission of both neutral and cationic silicon 
clusters under femtosecond laser ablation of silicon in vacuum. Cluster 
abundance and velocity distributions are found to depend strongly on laser 
fluence and the number of laser pulses applied. The obtained results provide 
clear evidence that different cluster formation mechanisms (Coulomb 
explosion, gas-phase condensation, phase explosion) are involved. Further 
work aimed at clarifying the contributions of these mechanisms for various 
ablation conditions is currently under way.

\begin{acknowledgments}
The work was supported by the International Science and Technology Center 
(Grant 2310).
\end{acknowledgments}

\end{document}